\documentclass[preprint,superscriptaddress,prl,fleqn]{revtex4} \newif\iftwocolumn \twocolumnfalse

\usepackage{amsmath}
\usepackage{amsfonts}
\usepackage{amssymb}
\usepackage{bm}
\usepackage{txfonts}
\usepackage[dvips]{graphicx}
\usepackage{multirow}
\usepackage{threeparttable}
\usepackage[usenames]{color}
\usepackage{ulem}

\def\bG{\bm G}
\def\bq{\bm q}

\def\br{\bm r}

\def\d{\partial }

\def\ep{\epsilon }

\def\s{\sigma }

\def\om{\omega }
\def\Om{\Omega }

\def\r{\mbox{\boldmath $r$} }
\def\rs{\mbox{\boldmath $\scriptstyle r$} }

\def\q{\mbox{\boldmath $q$} }

\def\Gs{\mbox{\boldmath $\scriptstyle G$} }

\def\GGamma{\mbox{\boldmath $\Gamma$} }
 
\def\AA{$\mathrm{\mathring{A}}$}

\def\bra#1{\langle\,{#1}\,\vert\,}
\def\ket#1{\,\vert\,{#1}\,\rangle}

\def\Bra#1{(\,{#1}\,\vert\,}
\def\Ket#1{\,\vert\,{#1}\,)}

\def\tilde{\widetilde}

\begin{document}

\preprint{Physical Review B, Rapid Communications (2016), to appear.}

\title{GW$\boldsymbol\Gamma$ $+$ Bethe--Salpeter equation approach for
photoabsorption spectra:
Importance of self-consistent GW$\boldsymbol\Gamma$ calculations in small atomic systems}


\author{Riichi Kuwahara}
\affiliation{
Department of Physics, Yokohama National University,
79-5 Tokiwadai, Hodogaya-ku, Yokohama 240-8501, Japan
}
\affiliation{
Dassault Syst\`emes BIOVIA K.K., ThinkPark Tower, 2-1-1 Osaki,
Shinagawa-ku, Tokyo 140-6020, Japan
}
\author{Yoshifumi Noguchi}
\affiliation{
Institute for Solid State Physics, The University of Tokyo,
5-1-5 Kashiwanoha, Kashiwa, Chiba 277-8581, Japan
}
\author{Kaoru Ohno}
\email[]{ohno@ynu.ac.jp}
\affiliation{
Department of Physics, Yokohama National University,
79-5 Tokiwadai, Hodogaya-ku, Yokohama 240-8501, Japan
}


\received{30 July 2015}
\revised{29 June 2016}

\begin{abstract}
The self-consistent GW$\Gamma$ method satisfies the Ward--Takahashi identity
(i.e., the gauge invariance or the local charge continuity)
for arbitrary energy ($\omega$) and momentum ($\bq$) transfers.
Its self-consistent first-principles treatment of the vertex $\Gamma=\Gamma_v$ or $\Gamma_W$ is possible
to first order in the bare ($v$) or dynamically-screened ($W$) Coulomb interaction. It
is developed
within a linearized scheme and combined with the Bethe--Salpeter equation (BSE) to accurately calculate
photoabsorption spectra (PAS) and photoemission (or inverse photoemission)
spectra (PES) simultaneously. The method greatly improves the PAS of
Na, Na$_3$, B$_2$, and C$_2$H$_2$ calculated using the standard one-shot
$G_0W_0$ $+$ BSE method that results in significantly
redshifted PAS by 0.8-3.1~eV, although the PES are well reproduced already in $G_0W_0$.
\end{abstract}


\maketitle



The quasiparticle (QP) equation method in many-body perturbation theory \cite{Hedin}
is powerful for simultaneously determining the
 photoemission (or inverse photoemission)
spectra (PES), i.e., QP energy spectra,
and QP wave functions of target materials from first-principles.
In this method, we expand the skeleton diagrams,
i.e., the diagrams drawn with the full Green's function lines, for the self-energy
in terms of the electron-electron Coulomb interaction $v$,
and solve the QP equation, which is equivalent to the Dyson equation,
as a self-consistent (SC) eigenvalue problem. The
Hartree--Fock (HF) approach provides the first-order approximation.
In Hedin's set of equations \cite{Hedin} known as the GW$\Gamma$ approach,
the exchange-correlation part of the self-energy
is expressed as $\Sigma_{\s}^{\rm xc}=iG_{\s}W\Gamma_{\s}$,
where $G_{\s}$ and $\Gamma_{\s}$ are the one-particle Green's function
and the vertex function ($\sigma$ is the spin index), respectively, and
$W=(1-vP)^{-1}v$
represents
the dynamically screened Coulomb interaction
($P=-i\sum_{\s}G_{\s}G_{\s}\Gamma_{\s}$ is the polarization function).
The simplest
approximation is to assume $\Gamma_{\s}=1$,
which is called the GW
approximation.

\begin{figure}[htbp]
\begin{center}
\includegraphics[width=80mm]{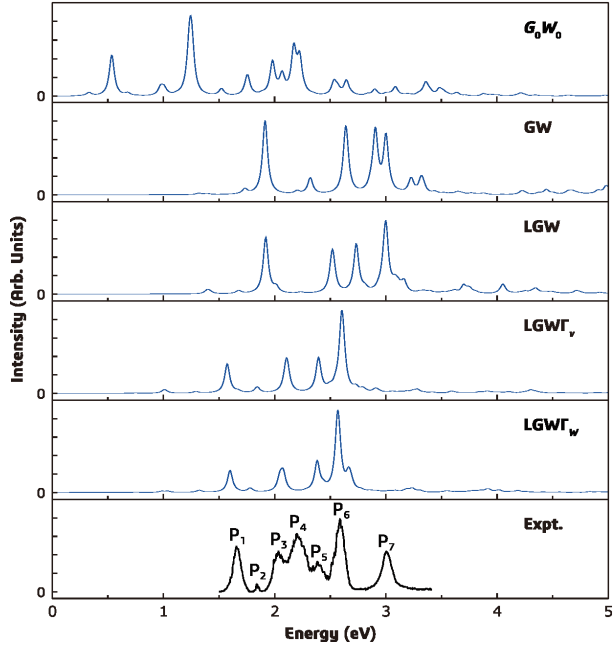}
\caption{Photoabsorption spectra of Na$_3$
calculated using $G_0W_0$, GW, LGW, LGW$\Gamma_v$, and LGW$\Gamma_W$.
Experimental data are taken from Ref. \cite{Wang}.}
\label{PAS}
\end{center}
\end{figure}

It is well known that 
the SC
GW
method
usually overestimates the energy gap \cite{Kotani,Shishkin},
while
the one-shot GW approach ($G_0W_0$) 
using the Kohn--Sham (KS) wave functions and eigenvalues \cite{HybertsenLouie}
results in a better energy gap.
However, quite recently, it has been pointed out that the photoabsorption spectra (PAS)
for small molecules
obtained by solving the Bethe--Salpeter equation (BSE) \cite{Onida,RohlfingLouie}
using
$G_0W_0$ are often significantly redshifted by about
1$\,$eV \cite{Hirose,Jacquemin}.
The use of the Heyd-Scuseria-Ernzerhof (HSE) functional or the 
SC GW calculation (hereafter referred to as GW)
improves the results, but they are not perfect \cite{Jacquemin,Bruneval}.
For a
spin-polarized sodium atom (Na) and trimer (Na$_3$),
$G_0W_0 +$ BSE
is
extremely bad,
although the $G_0W_0$ QP energies are reasonably good \cite{NoguchiIshiiOhnoSasaki}.
The calculated and experimental \cite{Na_Ex} optical gaps for Na
are
1.32$\,$eV and 2.10$\,$eV,
respectively,
and the calculated and experimental \cite{Wang} PAS for Na$_3$ are shown in
Fig. \ref{PAS}.
These calculated results are far off from the experimental data \cite{NoguchiOhno}.

Here,
we develop a 
GW$\Gamma$ method,
which involves 
a SC
treatment of the vertex $\Gamma=\Gamma_v$ or $\Gamma_W$ and satisfies
the Ward--Takahashi identity \cite{Ward,Takahashi,WT} to first order in $v$ or $W$,
and show that it remarkably improves the QP energies and the optical gaps of
spin-polarized
Na, Na$_3$, B$_2$, and closed-shell C$_2$H$_2$.
In this method, the 
SC one-particle Green's function,
i.e., 
SC
QP energies and 
wave functions
are obtained
in the GW$\Gamma$ scheme.
We
use the all-electron mixed basis approach, in which single particle wave functions
are expanded with both plane waves (PWs) and atomic orbitals (AOs) \cite{NoguchiIshiiOhnoSasaki,PRA}.
This Rapid Communication reports a 
first-principles
SC
GW$\Gamma$
calculation and its application to the BSE,
which has never been performed so far except for
some recent reports of 
non-SC
GW calculations
including the second-order screened exchange
by Ren {\it et al}. \cite{Ren}
and the GW$\Gamma^1$ method
(i.e., GW$^{\rm TC-TC}$ + single-shot vertex correction
for the self-energy with the static approximation)
by Gr\"{u}neis {\it et al}. \cite{Grueneis}.
All these authors used the KS, HF, or HSE wave functions throughout the calculations.
%

In the present SC GW$\Gamma$ $+$ BSE calculations, we will show the following:
(1)
 Highly reliable PES and PAS are simultaneously obtained for every system.
(2)
 All calculated results deviate by 0.1~eV at most from the available experimental data.
(3)
 The failure of the $G_0W_0$ $+$ BSE calculations for the
 PAS is caused by the use of localized KS wave functions above the vacuum level (VL),
and hence accurate QP wave functions are required.

Except for the $G_0W_0$ and GW calculations,
we use our recently developed technique \cite{PRA} to linearize the energy dependence of the self-energy $\Sigma_{\s}(\ep_n)$
to avoid the non-Hermitian problem caused by the energy dependence and to perform fully 
SC
calculations.
The important point of this technique is that $\Lambda_{\s} = \lim_{(\om,\bq)\rightarrow 0}\Gamma_{\s}(\br_1,\br_2,\bq;\mu+\om,\mu)
= 1 - \d\Sigma_{\s}(\om)/\d\om |_{\om=\mu}$ is the vertex function in the limit $(\om,\bq)\rightarrow 0$.
This is the Ward--Takahashi identity in the same limit.
[Here, $\mu=(\ep_{\rm HOMO}+\ep_{\rm LUMO})/2$.]
The QP equation is given by $H_{\s}\ket{n\s}=\ep_{n\s}\Lambda_{\s}\ket{n\s}$ with $H_{\s}=\hat{T}+\hat{v}_{\rm nuc}+\Sigma_{\s}(\mu)+\mu(\Lambda_{\s}-1)$.
Then, with the lower triangular matrix $L_{\s}$ in the Choleski decomposition \cite{Choleski} $\Lambda_{\s}=L_{\s}L^{\dagger}_{\s}$,
the renormalized QP states are given by $\ket{\tilde{n\s}}=L^{\dagger}_{\s}\ket{n\s}$, which satisfy
$\tilde{H}_{\s}\ket{\tilde{n\s}}=\ep_{n\s}\ket{\tilde{n\s}}$ with $\tilde{H}_{\s}=L^{-1}_{\s}H_{\s}L^{-1\dagger}_{\s}$
as well as the orthogonality and completeness conditions.
Moreover, the renormalized Green's function is given by $\tilde{G}_{\s}(\om)=L^{\dagger}_{\s}G_{\s}(\om)L_{\s}$; see
Ref. \cite{PRA}
for more details.

\begin{figure}[htbp]
\begin{center}
\includegraphics[width=80mm]{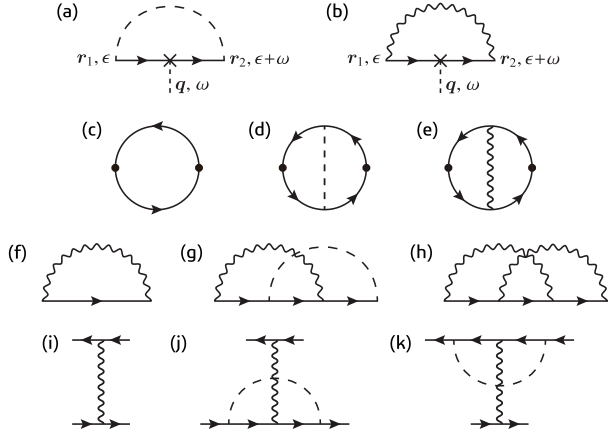}
\caption{Skeleton diagrams of the first-order vertex part (a) $\Gamma_{v\s}$ and (b) $\Gamma_{W\s}$;
(c), (d), and (e) are the polarization part $P$; (f), (g), and (h) are the exchange-correlation part of
the self-energy $\Sigma^{\rm xc}_{\s}$;
(i), (j), and (k) are the interaction kernel $\tilde{I}^{\;\s_1\s'_1}$ of
the BSE.
(c), (f), and (i) are usual diagrams without vertex correction;
(a), (d), (g), (j), and (k) involve the first-order vertex in $v$ (dotted line);
(b), (e), and (h) involve the first-order vertex in $W$ (wavy line).}
\label{diagram}
\end{center}
\end{figure}

({\sl Theorem 1})
In this linearized formulation, we can additionally introduce the 
vertex part
$\Gamma_{v\s}(\br_1,\br_2,\bq;\ep+\om,\ep)$ 
to first order in $v$
[Fig. \ref{diagram}(a)],
which we 
call the LGW$\Gamma_v$ method,
or
$\Gamma_{W\s}(\br_1,\br_2,\bq;\ep+\om,\ep)$
to first order
in $W$
[Fig. \ref{diagram}(b)],
which we 
call the LGW$\Gamma_W$ method.
These vertex parts depend fully on the energy and momentum transfers $\om$ and $\bq$, respectively,
at the center (cross in 
those figures). 
See the Supplemental Material (SM) \cite{SupplementalMaterial} for the proof of this theorem.

Then, the polarization function and the self-energy include 
the skeleton diagrams as shown in
Figs. \ref{diagram}(c)-\ref{diagram}(h).
{Figures \ref{diagram}(c) and \ref{diagram}(f) represent
the
diagrams
without a vertex; 
Figs. \ref{diagram}(d) and \ref{diagram}(g) and Figs. \ref{diagram}(e) and \ref{diagram}(h)
are the corresponding
vertex corrections to first order in $v$ ($\Gamma=$ $\Gamma_v$)
and $W$ ($\Gamma=$ $\Gamma_W$), respectively.
Figure \ref{Flow} illustrates the flow chart of the 
SC
LGW$\Gamma_W$ method.
The forms of the polarization function 
(Fig. \ref{diagram}(e)) and self-energy (Fig. \ref{diagram}(h))
are given in the SM.

({\sl Theorem 2}) The present LGW$\Gamma_v$ and LGW$\Gamma_W$  methods satisfy
the generalized Ward--Takahashi identity for arbitrary $\omega$ and $\q$,
which is equivalent to the gauge invariance (continuity equation for the electron density) \cite{Ward,Takahashi,WT},
up to first order in $v$ and $W$, 
respectively.
The proof is given in the SM. 



\begin{figure}[htbp]
\begin{center}
\includegraphics[width=80mm]{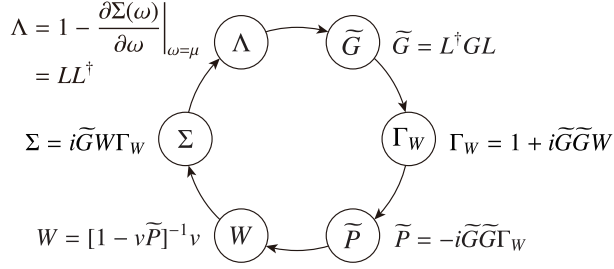}
\caption{Flow chart of the 
SC
LGW$\Gamma_W$ method.
$W$ and $\Gamma_W$ are replaced by $v$ and $\Gamma_v$
in the 
SC
LGW$\Gamma_v$ method.}
\label{Flow}
\end{center}
\end{figure}



Recently, the BSE
has been
solved
in
the one-shot second-order 
approach \cite{Zhang}.
In what follows,
we formulate the BSE for the LGW$\Gamma_v$ approach to spin-polarized
systems.
In the linearized formulation, we use the renormalized two-particle Green's function
	$\tilde{S}^{\sigma_1\sigma_2}_{\sigma'_1\sigma'_2}(x_1,x'_1;x_2,x'_2)
		= L_{\sigma'_1}L^{\dagger}_{\sigma_1}
		S^{\sigma_1\sigma_2}_{\sigma'_1\sigma'_2}(x_1,x'_1;x_2,x'_2)
		L_{\sigma_2}L^{\dagger}_{\sigma'_2}$
instead of 
$S^{\sigma_1\sigma_2}_{\sigma'_1\sigma'_2}(x_1,x'_1;x_2,x'_2)$,
defined by subtracting $\delta_{\sigma_1\sigma'_1}\delta_{\sigma_2\sigma'_2}G(x_1,x'_1)G(x_2,x'_2)$
from the original two-particle Green's function.
It satisfies the BSE
\begin{align}
	&\quad	
	\tilde{S}^{\sigma_1\sigma_2}_{\sigma'_1\sigma'_2}(x_1,x'_1;x_2,x'_2)
		= \tilde{G}_{\sigma_1}(x_1,x_2)\tilde{G}_{\sigma'_1}(x_2',x_1')
		\delta_{\sigma'_1\sigma'_2}\delta_{\sigma_1\sigma_2} \nonumber\\ 
	&\quad
	 + \sum_{\sigma_4\sigma'_4}\int \tilde{G}_{\sigma_1}(x_1,x_3)
		\frac{\delta\Sigma_{\sigma_3\sigma'_3}(x_3,x_3')}
		{\delta\tilde{G}_{\sigma_4\sigma'_4}(x_4,x_4')} \tilde{G}_{\sigma_1'}(x_3',x_1') 
		\; (L_{\sigma'_4}L^{\dagger}_{\sigma_4})^{-1}
		\nonumber \\
	&\quad 
	\times \;  
        	\tilde{S}^{\sigma_4\sigma_2}_{\sigma'_4\sigma'_2}(x_4,x_4';x_2,x_2') d^4x_3d^4x_3'd^4x_4d^4x_4',
\label{S=GG+GGXL^{-1}S}
\end{align}
where we used the fact that the original kernel
$\Xi^{\sigma_3\sigma_4}_{\sigma'_3\sigma'_4}(x_3,x_3';x_4,x_4')$ is
given by
\begin{align}
	\;\;\;\;
	\Xi^{\sigma_3\sigma_4}_{\sigma'_3\sigma'_4}
	=\frac{\delta\Sigma_{\sigma_3\sigma'_3}}{\delta G_{\sigma_4\sigma'_4}}
	=\frac{\delta\tilde{G}_{\sigma_3\sigma'_3}}{\delta G_{\sigma_3\sigma'_3}}
	\frac{\delta\Sigma_{\sigma_3\sigma'_3}}{\delta\tilde{G}_{\sigma_4\sigma'_4}}
	= L_{\sigma_3}\frac{\delta\Sigma_{\sigma_3\sigma'_3}}
	{\delta\tilde{G}_{\sigma_4\sigma'_4}}L^{\dagger}_{\sigma'_3}.
\label{Xi}
\end{align}
The functional derivative $\delta\Sigma_{\sigma_3\sigma'_3}(x_3,x_3')
/\delta\tilde{G}_{\sigma_4\sigma'_4}(x_4,x_4')$ is given by
$
 	-i\delta_{\sigma_3\sigma'_3}\delta_{\sigma_4\sigma'_4}
		\delta(x_3-x_3')\delta(x_4-x_4')
	v(\r_3-\r_4)
	+ \delta_{\sigma_3\sigma_4} \delta_{\sigma'_3\sigma'_4}
	\tilde{I}
	^{\sigma_3\sigma'_3}
	(x_3,x_3';x_4,x_4').
$
Ignoring all terms having functional derivatives of $W[\tilde{G}]$ 
with respect to $\tilde{G}$ as usual \cite{Strinati},
we have $\tilde{I}
^{\sigma_3\sigma'_3}
(x_3,x_3';x_4,x_4')$ expressed as
$i
\{ \delta(x_3-x_4)\delta(x_3'-x_4')W(x_3,x_3')
+ \delta(x_3-x_4)[W\Gamma_{v\sigma'_3}](x_3',x_4';x_3)
+ \delta(x_3'-x_4')[W\Gamma_{v\sigma_3}](x_3,x_4;x_3') \}$,
which is represented by the skeleton diagrams of
Figs. \ref{diagram}(i), \ref{diagram}(j), and \ref{diagram}(k).
Here the last two terms
(Figs. \ref{diagram}(j) and \ref{diagram}(k))
come from vertex correction to first order in $v$.
From these equations,
we find that $\Lambda_{\sigma}=L_{\sigma}L^{\dagger}_{\sigma}$
should be multiplied to the polarization function as
$\tilde{P}^{\;\sigma}_{\Gs\Gs'}=P^{\;\sigma}_{\Gs\Gs'}\Lambda_{\sigma}$ \cite{PRA}.
Then, putting
$V_{\nu\mu dc}^{\s_1\s_2} = \sum_{\Gs} \bra{\tilde{\nu\s_1}} e^{i\Gs\cdot\rs}
\ket{\tilde{\mu\s_1}} \bra{d\s_2} e^{-i\Gs\cdot\rs'} \ket{c\s_2} v(\bG)$
and using the expression for
$\tilde{I}^{\;\s_1\s'_1}_{\nu\mu dc}(\om)$
given in the SM, 
we obtain the matrix eigenvalue equation of the BSE \cite{Strinati}
\begin{gather}
\qquad
(\epsilon_{c'\sigma_1}-\epsilon_{d'\sigma'_1}-\Om_r)A^r_{d'\sigma'_1,c'\sigma_1}
\iftwocolumn \nonumber\\ \fi
= - \sum_d^{\rm occ}\sum_c^{\rm emp}\left\{ \delta_{\sigma_1\sigma'_1}\sum_{\sigma_2}
V^{\sigma_1\sigma_2}_{c'd'dc} A^r_{d\sigma_2,c\sigma_2}
 - \tilde{I}^{\;\sigma_1\sigma'_1}_{c'd'dc}(\Om_r) A^r_{d\sigma'_1,c\sigma_1}\right\}
\label{eigenEq}
\end{gather}
in the Tamm--Dancoff approximation \cite{Strinati}.
We also use this formulation in the LGW$\Gamma_W$ approach,
because 
the resulting error is on the order of 0.01~eV.

For spin-polarized systems,
we have to generally solve the eigenvalue equation
(\ref{eigenEq})
in the $\uparrow\uparrow$-$\downarrow\downarrow$ subspace,
\begin{eqnarray}
\qquad
\left(\hskip-1mm
\begin{array}{cc}
 h^{\uparrow\uparrow}+v^{\uparrow\uparrow} &  v^{\uparrow\downarrow} \\
 v^{\downarrow\uparrow} & h^{\downarrow\downarrow}+v^{\downarrow\downarrow} \\
\end{array}
\hskip-1mm\right)
\left(\hskip-1mm
\begin{array}{c}
 A^r_{\uparrow\uparrow} \\
 A^r_{\downarrow\downarrow} \\
\end{array}
\hskip-1mm\right) = 0,
\label{2x2}
\end{eqnarray}
where we put $h^{\sigma_1\sigma'_1}=
(\epsilon_{c'\sigma_1}-\epsilon_{d'\sigma'_1}-\Om_r)\delta_{cc'}\delta_{dd'}
 - \tilde{I}^{\;\sigma_1\sigma'_1}_{c'd'dc}$
and
$v^{\sigma_1\sigma_2}=V^{\sigma_1\sigma_2}_{c'd'dc}$.

We used a face-centered cubic unit cell with edge length of
14$\,$\AA{ }for Na and B$_2$, 15$\,$\AA{ }for C$_2$H$_2$, and 18$\,$\AA{ }for Na$_3$.  
All of the core and (truncated) valence numerical AOs are used together with the PWs.
The atomic geometries are optimized with DMol$^3$ \cite{DMol3a,DMol3b}. 
The bond lengths are
3.23$\,$\AA, 3.23$\,$\AA, and 5.01$\,$\AA{ }for Na$_3$, 1.61$\,$\AA{ }for B$_2$, 1.20$\,$\AA{ }for C$\equiv$C, and 1.06$\,$\AA{ }for
C-H
at the Becke three-parameter Lee-Yang-Parr (B3LYP) functional level.
We used
3.61$\,$(50.76)$\,$Ry, 1.23$\,$(30.7)$\,$Ry, 6.82$\,$(38.1)$\,$Ry, and 11.1$\,$(44.2)$\,$Ry
cutoff energies for PWs
(for $\Sigma^{\rm x}_{\s}$), respectively, for Na, Na$_3$, B$_2$, and C$_2$H$_2$.
The cutoff energy for $P$ and $\Sigma^{\rm c}_{\s}$ is the same as that for PWs for Na and Na$_3$,
and is set at 4.57$\,$Ry for B$_2$ and 3.98$\,$Ry for C$_2$H$_2$.
We used the full $\om$-integration \cite{Omega} and the projection operator for the GW-related calculations,
but used the plasmon-pole model \cite{Linden-Horsch} and 600 empty states for the $\Gamma$-related calculations
as well as for solving the BSE in order to save the computational cost.

The 
resulting ionization potential (IP), electron affinity (EA), and
optical gap $E^{\rm opt}_g$ (corresponding to the first dipole-allowed transition)
of Na, Na$_3$, B$_2$, and C$_2$H$_2$
calculated using the $G_0W_0$, GW, LGW, 
and LGW$\Gamma_W$ methods
are listed in Tables \ref{NaNa3} and \ref{B2C2H2}, together with
the results of previous multireference single and double configuration interaction
(MRDCI) calculations \cite{Na_CI_IP,NaNa3_CI_EA_OPT,Na3_CI_IP,B2_CI_IP,B2_CI_EA,B2_CI_Opt,acetylene_CI_IP},
configuration interaction single and double (CID) calculations \cite{acetylene_CI_Opt},
and the corresponding experimental
values \cite{Na_Ex,Wang,NaNa2_IP,Na_EA,Na3_IP,Na3_EA,B2_IP,B2_EA,B2_Opt,B2_Opt2,acetylene_IP,acetylene_Opt}.
For Na and Na$_3$, the results of LGW$\Gamma_v$ are also listed in Table \ref{NaNa3}.
Note that EA of C$_2$H$_2$ is negative and not shown in Table \ref{B2C2H2}.
Let us first compare the results of IP and EA with the experimental values.
$G_0W_0$ results in reasonable IP and EA (IP of C$_2$H$_2$ is similar to those obtained in Ref. \cite{GW100})
while GW has a tendency to overestimate IP and underestimate EA,
although the experimental error bar is large for B$_2$.
LGW improves GW \cite{PRA}, but is not perfect.
In contrast, LGW$\Gamma_v$ and LGW$\Gamma_W$ almost perfectly improve both IP and EA.
For LGW$\Gamma_W$, the deviation from the experimental values is
0.03$\,$eV
for Na,
0.07$\,$eV
for Na$_3$,
and 0.01~eV for C$_2$H$_2$.
Compared with previous MRDCI calculations \cite{Na_CI_IP,NaNa3_CI_EA_OPT,Na3_CI_IP,B2_CI_IP,B2_CI_EA,acetylene_CI_IP},
our results are closer to the experimental IP and EA for almost all cases.


\begin{table}[htbp]
\begin{center}
\caption{Ionization potential (IP), electron affinity (EA), and optical gap
$E^{\rm opt}_g$ (corresponding to ${}^2$S $\rightarrow$ ${}^2$P
and ${}^2$B$_2$ $\rightarrow$ ${}^2$A$_1$ transitions) of Na and Na$_3$ (in units of eV).}
\begin{threeparttable}
{\tabcolsep = 2.5mm
\begin{tabular}{cccccccc}\hline\hline
        	& \multicolumn{3}{c}{Na}	&& \multicolumn{3}{c}{Na$_3$} \\ \cline{2-4} \cline{6-8}
        	& IP	& EA	& $E^{\rm opt}_g$	&& IP	& EA	& $E^{\rm opt}_g$	\\ \hline
$G_0W_0$	& 5.15	& 0.41	& 1.32	&& 4.10	& 1.14 & 0.53	\\ \hline
GW			& 5.40	& 0.33	& 2.23	&& 4.64	& 0.51 & 1.91	\\ \hline
LGW			& 5.23	& 0.42	& 2.18	&& 4.48 & 0.66 & 1.92	\\ \hline
LGW$\Gamma_v$ & 5.01& 0.60	& 2.00	&& 4.08 & 1.04 & 1.57	\\ \hline
LGW$\Gamma_W$ & 5.12& 0.58	& 2.16	&& 4.04 & 1.15 & 1.60	\\ \hline
MRDCI		& 4.97\tnote{a}	& 0.44\tnote{b}	& 1.98\tnote{b} && 3.76\tnote{c} & 1.07/1.17\tnote{b} & 1.61\tnote{b} \\ \hline 
Expt.		& 5.14\tnote{d}	& 0.55\tnote{e}	& 2.10\tnote{f} && 3.97\tnote{g}	& 1.02/1.16\tnote{h} & 1.65\tnote{i} \\ \hline
\hline
\end{tabular}
}
\begin{tablenotes}\scriptsize
\item[a] Reference \cite{Na_CI_IP}. 
\item[b]
Reference \cite{NaNa3_CI_EA_OPT}. 
\item[c]
Reference \cite{Na3_CI_IP}. 
\item[d]
Reference \cite{NaNa2_IP}. 
\item[e]
Reference \cite{Na_EA}. 
\item[f]
Reference \cite{Na_Ex}. 
\item[g]
Reference \cite{Na3_IP}. 
\item[h]
Reference \cite{Na3_EA}. 
\item[i]
Reference \cite{Wang}.
\end{tablenotes}
\end{threeparttable}
\label{NaNa3}
\end{center}
\end{table}

\begin{table}[htbp]
\begin{center}
\caption{IP, EA, and $E^{\rm opt}_g$ (corresponding to the ${}^3\Sigma^-_g\rightarrow{}^3\Sigma^-_u$ transition) of B$_2$,
and IP and $E^{\rm opt}_g$ (corresponding to the ${}^1\Sigma^+_g\rightarrow{}^1\Pi_u$ transition) of C$_2$H$_2$
(in units of eV).}
\begin{threeparttable}
{\tabcolsep = 2mm
\begin{tabular}{ccccccc}\hline\hline
        	& \multicolumn{3}{c}{B$_2$}	&& \multicolumn{2}{c}{C$_2$H$_2$} \\ \cline{2-4}\cline{6-7}
        		& IP 				& EA 		& $E^{\rm opt}_g$ && IP	& $E^{\rm opt}_g$ \\ \hline
$G_0W_0$		& 9.21				& 2.18		& 2.44	&& 11.05	& 5.01		\\ \hline
GW				& 9.97				& 1.76		& 3.94	&& 11.65	& 8.39	\\ \hline
LGW				& 9.79				& 1.94		& 3.75	&& 11.44	& 8.23	\\ \hline
LGW$\Gamma_W$	& 9.87	& 1.91		& 3.84	&& 11.48	& 8.25	\\ \hline
MRDCI			& 9.48\tnote{a}	& 2.0\tnote{b}	& 3.85\tnote{c} && 11.21\tnote{d} & (8.06)\tnote{e} \\ \hline 
Expt.			& 10.3$\pm 0.6$\tnote{f}& 1.8$\pm 0.4$\tnote{g}	& 3.79\tnote{h} && 11.49\tnote{i} & 8.16\tnote{j}	\\
\hline\hline 
\end{tabular}
}
\begin{tablenotes}\scriptsize
\item[a]
Reference \cite{B2_CI_IP}. 
\item[b]
Reference \cite{B2_CI_EA}. 
\item[c]
CASSCF/MRDCI:~Reference \cite{B2_CI_Opt}. 
\item[d]
Reference \cite{acetylene_CI_IP}. 
\item[e]
CID:~Reference \cite{acetylene_CI_Opt}. 
\item[f]
Reference \cite{B2_IP}. 
\item[g]
Reference \cite{B2_EA}. 
\item[h]
Reference \cite{B2_Opt,B2_Opt2}. 
\item[i]
Reference \cite{acetylene_IP}. 
\item[j]
Reference \cite{acetylene_Opt}. 
\end{tablenotes}
\end{threeparttable}
\label{B2C2H2}
\end{center}
\end{table}


Next we compare the results of the optical gap $E^{\rm opt}_g$ with experiments.
$G_0W_0$ significantly underestimates the experimental $E^{\rm opt}_g$ for all systems and
GW overestimates the experimental $E^{\rm opt}_g$.
The deviation from the experimental values is
0.8-3.1$\,$eV for $G_0W_0$ and
0.13-0.26$\,$eV for GW.
LGW improves the results except for Na$_3$;
the deviation from the experimental values is
0.08$\,$eV
for Na,
0.27$\,$eV
for Na$_3$,
0.04$\,$eV
for B$_2$, and
0.07$\,$eV
for C$_2$H$_2$.
In contrast, LGW$\Gamma_v$ and LGW$\Gamma_W$ give excellent $E^{\rm opt}_g$ for all systems.
For LGW$\Gamma_W$, the difference between the theoretical and experimental values is less than
0.06$\,$eV for Na and Na$_3$, 0.05$\,$eV for B$_2$, and 0.09$\,$eV for C$_2$H$_2$.
Compared with the experimental values,
our $E_g^{\rm opt}$ is better than the previous MRDCI results for Na \cite{NaNa3_CI_EA_OPT},
and CID results for C$_2$H$_2$ \cite{acetylene_CI_Opt}, or comparable to (differs only by
0.01$\,$eV
from) previous MRDCI results for Na$_3$ \cite{NaNa3_CI_EA_OPT},
and complete-active-space self-consistent-field (CASSCF)/MRDCI results for B$_2$ \cite{B2_CI_Opt}.
The LGW$\Gamma_W$ $+$ BSE photoabsorption peak, i.e., the exciton wave function mainly consists of
the following QP hole and electron level pair(s):
6($s\hskip-1mm\uparrow$) $\rightarrow$ 7($p\hskip-1mm\uparrow$) for Na, 17($s\hskip-1mm\uparrow$) $\rightarrow$ 19($p\hskip-1mm\uparrow$) for Na$_3$, 4($\s\hskip-1mm\uparrow$) $\rightarrow$ 7($\pi\hskip-1mm\uparrow$) and 3($\s\hskip-1mm\downarrow$) $\rightarrow$ 6($\pi\hskip-1mm\downarrow$) for B$_2$, and 7($\pi$) $\rightarrow$ 8($\s$) for C$_2$H$_2$.
Figure
\ref{PAS} shows the PAS of Na$_3$ calculated using $G_0W_0$,
GW, LGW, LGW$\Gamma_v$, and LGW$\Gamma_W$
and the experimental spectra \cite{Wang}.
The
overall spectral shapes are similar in all these methods
except for $G_0W_0$,
although the peak positions are almost constantly shifted by an amount
indicated by the difference between the calculated and experimental $E^{\rm opt}_g$'s in Table \ref{NaNa3},
and the peak heights somewhat change after the inclusion of the vertex correction.
Obviously, GW and LGW overestimate the peak positions,
while LGW$\Gamma_v$ and LGW$\Gamma_W$ give good peak positions except for P$_4$ and P$_7$.
(LGW$\Gamma_v$ has a small peak at 2.9~eV, which may correspond to P$_7$.)
The remaining discrepancy between the theory and experiment
in the case of Na$_3$
may be
mainly
attributed to
the neglect
of isomers and the atomic vibration effects.

Figure
\ref{EnergyLevel} shows the QP (or KS) energy spectra calculated
using the local density approximation (LDA), $G_0W_0$, and LGW$\Gamma_W$.
The experimental IP and EA are indicated by IP and EA on the right vertical border line.
Now we 
discuss
the reason why
the PAS calculated using $G_0W_0 +$ BSE
are
so poor.
For Na$_3$, the number of up-spin (down-spin) levels below 
the VL
is 26$\,$(26) for LDA, 20$\,$(19) for GW, 20$\,$(20) for LGW, and 22$\,$(20) for LGW$\Gamma_v$ and LGW$\Gamma_W$.
We confirmed that the KS and QP wave functions very much resemble each other
for the first 20 
levels below 
the VL.
However, they are quite different for the QP levels above 
the VL.
For example, the spin-up and spin-down KS wave functions at the
21st down-spin level and the 23rd up-spin 
level below 
the VL
are depicted in
Fig. \ref{EnergyLevel}.
They
are
localized.
However, the corresponding $G_0W_0$ QP energies are both above 
the VL
and the full QP wave functions
are not
localized.
In our $G_0W_0$ + BSE calculation of Na$_3$, the first small photoabsorption peak around
0.35$\,$eV
 (see the top panel in
Fig. \ref{PAS})
mainly consists of
the QP hole and electron level pairs between 16 $\rightarrow$ 17$\,$(19.8\%), 21$\,$(8.7\%) for down-spin,
and 17 $\rightarrow$ 18$\,$(21.8\%), 20$\,$(31.4\%), 21$\,$(9.0\%), 23$\,$(2.3\%) for up-spin.
The unphysically bound KS wave functions of the 21$\downarrow$ and 23$\uparrow$ levels
contribute to the BSE matrix elements, leading to
unphysically
large
electron-hole interactions
and in turn, to
the optical transitions with
very
small photoabsorption energies.
This gives the wrong spectra for $G_0W_0$ in
Fig. \ref{PAS}.
It
has already been known for more than 50 years \cite{Baym-Kadanoff}
that the BSE should be solved with the fully 
SC
Green's function
in order to satisfy the conservation laws as well as the longitudinal $f$-sum rule.
However, the QP gap and the optical gap obtained using the 
GW method are
blueshifted because they do not satisfy the generalized Ward--Takahashi
identity.
To improve the result, it is necessary to use the 
GW$\Gamma$ method.

\begin{figure}[htbp]
\begin{center}
\includegraphics[width=80mm]{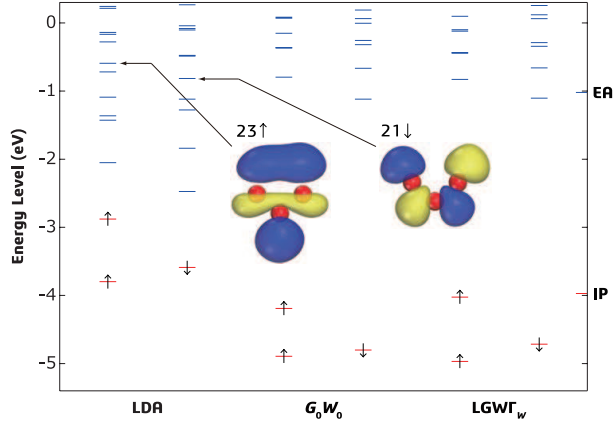}
\caption{QP (or KS) energy spectra of Na$_3$ calculated using the
LDA, $G_0W_0$, and LGW$\Gamma_W$. Unphysically bound KS wave functions
at the 21st spin-down level and the 23rd spin-up level
are also depicted.
Red balls are Na atoms, while yellow and blue colors
indicate the positive and negative regions
of the wave function, respectively.}
\label{EnergyLevel}
\end{center}
\end{figure}

In this
Rapid Communication,
we presented the 
$G_0W_0$,
GW, LGW,
and LGW$\Gamma_W$ 
(LGW$\Gamma_v$)
calculations for Na, Na$_3$, B$_2$, and C$_2$H$_2$.
If the $G_0W_0$ QP energies are used together with the KS wave functions,
there is inconsistency between the QP energies and wave functions
at some levels
above 
the VL.
Moreover, the 
GW and LGW methods are not sufficient
because they overestimate both the QP energy gap and optical gap.
To obtain better gap estimates, it is necessary
to perform the 
GW$\Gamma$ calculation.
We showed that the 
LGW$\Gamma_W$ method produces consistent and the best
PES
and PAS among all of the methods used in this study.
The self-consistent treatment of $\Gamma$ is required to obtain consistently good results for both PES and PAS,
and its computational cost scales as O($N^2M^3$), where
$N$ and $M$
are the numbers of basis functions and empty states, respectively,
if we use the plasmon-pole model.
The present method is applicable to vertical transitions but
cannot handle relaxation processes.

\begin{acknowledgments}
This work was supported by the Grant-in-Aid for Scientific Research
B (Grant No. 25289218) from JSPS and also by the Grant-in-Aid
for Scientific Research on Innovative Areas (Grant No. 25104713) from MEXT.
We are also indebted to the HPCI promoted by MEXT
for the use of the supercomputer SR16000 at Hokkaido University
and at IMR, Tohoku University (Project IDs. hp140214, hp150231, hp160072, and hp160234).
\end{acknowledgments}


\begin{thebibliography}{27}%
\bibitem{Hedin}%
L. Hedin, Phys. Rev. {\bf 139}, A796 (1965).
\bibitem{Kotani}
T. Kotani, M. van Schilfgaarde, and S. V. Faleev, Phys. Rev. B {\bf 76}, 165106 (2007).
\bibitem{Shishkin}%
M. Shishkin, M. Marsman, and G. Kresse,
Phys. Rev. Lett. {\bf 99}, 246403 (2007).
\bibitem{HybertsenLouie}%
M. S. Hybertsen and S.~G. Louie, Phys. Rev. Lett. {\bf 55}, 1418 (1985).
\bibitem{Onida}
G. Onida, L. Reining, R. W. Godby, R. Del Sole, and W. Andreoni,
Phys. Rev. Lett. {\bf 75}, 818 (1995).
\bibitem{RohlfingLouie}
M. Rohlfing and S.~G. Louie,
Phys. Rev. Lett. {\bf 80}, 3320 (1998).
\bibitem{Hirose}
D. Hirose, Y. Noguchi, and O. Sugino, Phys. Rev. B {\bf 91}, 205111 (2015).
\bibitem{Jacquemin}
D. Jacquemin, I. Duchemin, X. Blase, J. Chem. Theory Comput. {\bf  11}, 3290 (2015).
\bibitem{Bruneval}
F. Bruneval, S. M. Hamed, and J. B. Neaton, J. Chem. Phys. {\bf 142}, 244101 (2015).
\bibitem{NoguchiIshiiOhnoSasaki}
Y. Noguchi, S. Ishii, K. Ohno, and T. Sasaki, J. Chem. Phys. {\bf 129}, 104104 (2008).
\bibitem{Na_Ex}%
J.~E. Sansonetti, J. Phys. Chem. Ref. Data {\bf 37}, 1659 (2008).
\bibitem{Wang}
C. R. C. Wang, S. Pollack, T. A. Dahlseid, G. M. Koretsky, and M. M. Kappes,
J. Chem. Phys. {\bf 96}, 7931 (1992).
\bibitem{NoguchiOhno}
For the even number clusters of sodium, the $G_0W_0 +$ BSE result is not so bad;
see Ref. \cite{Onida} and: Y. Noguchi and K. Ohno, Phys. Rev. A {\bf 81}, 045201 (2010).
\bibitem{Ward}
J. C. Ward, Phys. Rev. {\bf 78}, 182 (1950);
\bibitem{Takahashi}
Y. Takahashi, Il Nuovo Cimento {\bf 6}, 371 (1957).
\bibitem{WT}
J. R. Schrieffer, {\it Theory of Superconductivity}
(Advanced Book Classic) (Westview Press, 1999) Chap. 8-5.
\bibitem{PRA}
R. Kuwahara and K. Ohno, Phys. Rev. A. {\bf 90}, 032506 (2014).
\bibitem{Ren}
X. Ren, P. Rinke, G. E. Scuseria, and M. Scheffler,
Phys. Rev. B {\bf 88}, 035120 (2013).
\bibitem{Grueneis}
A. Gr\"{u}neis, G. Kresse, Y. Hinuma, and F. Oba,
Phys. Rev. Lett. {\bf 112}, 096401 (2014).
\bibitem{Choleski} It is also possible to use $L=\Lambda^{1/2}$ instead of
the Choleski decomposition in the linearization procedure as in Ref. \cite{Shishkin}.
\bibitem{SupplementalMaterial} 
See Supplemental Material at http://link.aps.org/supplemental/10.1103/PhysRevB.xx.xxxxxx
for the proof of Theorem 1 and Theorem 2.
\bibitem{Zhang}
D. Zhang, S. N. Steinmann, and W. Yang,
J. Chem. Phys. {\bf 139}, 154109 (2013).
\bibitem{Strinati}
G. Strinati,
Phys. Rev. Lett. {\bf 49}, 1519 (1982).
\bibitem{DMol3a}
B. Delley, J. Chem. Phys. {\bf 92},  508  (1990).
\bibitem{DMol3b}
B. Delley, J. Chem. Phys. {\bf 113}, 7756 (2000).
\bibitem{Omega}
M. Zhang, S. Ono, N. Nagatsuka, and K. Ohno, Phys. Rev. B {\bf 93}, 155116 (2016).
\bibitem{Linden-Horsch}%
W. von~der Linden and P. Horsch, Phys. Rev. B {\bf 37}, 8351 (1988).
\bibitem{Na_CI_IP} R. J. Buenker and S. Krebs, in {\it Recent Advances in Multireference Methods},
edited by K. Hirao,
Recent Advances in Computational Chemistry Vol. 4 (World Scientific, Singapore, 1999) pp.1-29.
\bibitem{NaNa3_CI_EA_OPT} V. Bona\v{c}i\'{c}-Kouteck\'{y}, P. Fantucci, and J. Kouteck\'{y},
 J. Chem. Phys. {\bf 93}, 3802 (1990).
\bibitem{Na3_CI_IP} V. Bona\v{c}i\'{c}-Kouteck\'{y}, I. Boustani, M. Guest, and J. Kouteck\'{y},
 J. Chem. Phys. {\bf 89}, 4861 (1988).
\bibitem{B2_CI_IP} P. J. Bruna and J. S. Wright, J. Phys. Chem. {\bf 94}, 1774 (1990).
\bibitem{B2_CI_EA} P. J. Bruna and J. S. Wright, J. Phys. B {\bf 23}, 2197S (1990).
\bibitem{B2_CI_Opt} S. R. Langhoff and C. W. Bauschlicher, Jr., J. Chem. Phys. {\bf 95}, 5882 (1991).
\bibitem{acetylene_CI_IP} W. P. Kraemer and W. Koch, Chem. Phys. Lett. {\bf 212}, 631 (1993).
\bibitem{acetylene_CI_Opt} W. E. Kammer, Chem. Phys. {\bf 5}, 408 (1974).
\bibitem{NaNa2_IP}%
A. Herrmann, E. Schumacher, and L. W\"oste, J Chem. Phys. {\bf 68}, 2327 (1978).
\bibitem{Na_EA}%
H. Hotop and W.~C. Lineberger, J. Phys. Chem. Ref. Data {\bf 14},
731 (1985).
\bibitem{Na3_IP}
A. Herrmann, S. Leutwyler, E. Schumacher, and L. Woste, Helv. Chim. Acta {\bf 61}, 453 (1978).
\bibitem{Na3_EA}
K.~M. McHugh, J.~G. Eaton, G.~H. Lee, H.~W. Sarkas, L.~H. Kidder, J.~T. Snodgrass, M.~R. Manaa, and K.~H. Bowen,
J. Chem Phys. {\bf 91}, 3792 (1989).
\bibitem{B2_IP} L. Hanley, J. L. Whitten, and S. L. Anderson,
 J. Phys. Chem. {\bf 92}, 5803 (1988).
\bibitem{B2_EA} C. J. Reid, Int. J. Mass Spect. Ion Proces. {\bf 127}, 147 (1993).
\bibitem{B2_Opt} W. R. M. Graham and W. Weltner Jr.,
J. Chem. Phys. {\bf 65}, 1516 (1976).
\bibitem{B2_Opt2} K. P. Huber and G. Herzberg, {\it Molecular Spectra and Molecular Structure: IV. Constants of Diatomic Molecules} (Van Nostrand Reinhold Company, New York, 1979).
\bibitem{acetylene_IP}
G. Bieri and L. {\AA}sbrink, J. Elec. Spec. Rel. Phenom. {\bf 20}, 149 (1980).
\bibitem{acetylene_Opt} W. C. Price, Phys. Rev. {\bf 47}, 444 (1935).
\bibitem{GW100} M. J. van Setten, F. Caruso, S. Sharifzadeh, X. Ren, M. Scheffler,
F. Liu, J. Lischner, L. Lin, J. R. Deslippe, S. G. Louie, C. Yang, F. Weigend, J. B. Neaton,
F. Evers, and P. Rinke, J. Chem. Theory Comput. {\bf 11}, 5665 (2015).
\bibitem{Baym-Kadanoff}
G. Baym and L.~P. Kadanoff, Phys. Rev. {\bf 124}, 287 (1961).
\end{thebibliography}

%

\newpage

{\Large Supplemental Material}

\vskip5mm

\begin{center}

 * * * * * *

{\large GW$\boldsymbol\Gamma$ $+$ Bethe--Salpeter equation approach for photoabsorption spectra:

Importance of self-consistent GW$\boldsymbol\Gamma$ calculations in small atomic systems}

* * * * * * *

Riichi Kuwahara,$^{1,2}$ Yoshifumi Noguchi,$^3$ and Kaoru Ohno$^{1,*}$

$^{\sl 1}$ {\sl Department of Physics, Yokohama National University,

79-5 Tokiwadai, Hodogaya-ku, Yokohama 240-8501, Japan}

$^{\sl 2}$ {\sl Dassault Syst\`emes BIOVIA K.K., ThinkPark Tower,

2-1-1 Osaki, Shinagawa-ku, Tokyo 140-6020, Japan}

$^{\sl 3}$ {\sl Institute for Solid State Physics, The University of Tokyo,

5-1-5 Kashiwanoha, Kashiwa, Chiba 277-8581, Japan}

\end{center}

\vspace*{1cm}

\def\theequation{S.\arabic{equation}}

In this Supplemental Material, we prove Theorem 1 and Theorem 2,
and present explicit forms of the polarization function, the self-energy,
and the interaction kernel $\tilde{I}^{\;\s_1\s'_1}_{\nu\mu dc}(\om)$.
Figure and reference numbers refer to the figures and references in the Letter
except for Ref [S1]. 

\begin{itemize}
\item[Theorem 1.] In the linearized formulation, one can additionally introduce the vertex part
$\Gamma_{v \s}(\br_1,\br_2,\bq;\ep+\om,\ep)$
at the first order in the bare Coulomb interaction $v$ (Fig. 2(a)), which we will call the LGW$\Gamma_v$ method,
or
$\Gamma_{W \s}(\br_1,\br_2,\bq;\ep+\om,\ep)$ at the first order
in the dynamically screened Coulomb interaction $W$ (Fig. 2(b)), which we will call the LGW$\Gamma_W$ method.
\item[Theorem 2.] The LGW$\Gamma_v$ and LGW$\Gamma_W$ methods satisfy
the Ward--Takahashi identity to the first order in the bare Coulomb interaction $v$
and in the dynamically screened Coulomb interaction $W$, respectively.
\end{itemize}

\newpage

\section{Proof of Theorem 1}

The statement of Theorem 1 holds because using $\tilde{G}_{\s}(\om)$ in place of $G_{\s}(\om)$ in the linearized formulation
introduces the $\om=\bq=0$ vertex function $\Lambda_{\s}$ just at the other side of the interaction line
where we introduced the first-order vertex part $\Gamma_{v \s}(\br_1,\br_2,\bq;\ep+\om,\ep)$
or $\Gamma_{W \s}(\br_1,\br_2,\bq;\ep+\om,\ep)$ for arbitrary $\om$ and $\bq$.
Therefore there is no double counting in the vertex correction up to the first order in $v$ or $W$.
In this way, the LGW$\Gamma_v$ and LGW$\Gamma_W$ methods rigorously treat the vertex parts
to the first order in $v$ (Fig. 2(a)) and in $W$ (Fig. 2(b)), respectively.
These vertex parts depend fully on the energy and momentum transfers $\om$ and $\bq$
at the center (cross) as well as the frequencies and the coordinates at both ends; see Figs. 2(a) and (b).

Moreover, in the LGW$\Gamma_v$ method, it is possible to show that there is no interference between
the first-order vertex part of Fig. 2(a) (i.e., the vertex part at the first order in $v$)
and $\Lambda_{\s}$ in the $\om=\bq=0$ limit.
In the $\om=\bq=0$ limit, the former can be expressed as the $-\om$ derivative of the $\om$-independent Fock exchange
self-energy $\Sigma^{\rm x}_{\sigma}$ and hence exactly equals to zero, i.e.,
$\Gamma_{v \s}(\br_1,\br_2,\bq=0;\ep,\ep) = -\d\Sigma^{\rm x}_{\sigma}/\d\om=0$.
Therefore, the full vertex part $\Lambda_{\s}$ in the $\om=\bq=0$ limit
introduced in the linearized formulation does not interfere to the first-order vertex part of Fig. 2(a).
In the LGW$\Gamma_W$ method, however, the first-order vertex part of Fig. 2(b)
(i.e., the vertex part at the first order in $W$) may interfere  with the full vertex part $\Lambda_{\s}$ in the $\om=\bq=0$ limit
at higher orders beyond the present approximation.

\section{Proof of Theorem 2}

Ward--Takahashi identity [16-18] is given by
\begin{eqnarray}
&& \delta^4(x_1-x)G^{-1}(x,x_2) - G^{-1}(x_1,x)\delta^4(x-x_2)
\nonumber\\
&=& i\nabla\cdot\GGamma(x_1,x_2,x) + i\frac{\d}{\d t}\Gamma(x_1,x_2,x).
\end{eqnarray}
If we Fourier transform this equation with respect to $t-t_2$
and $t_1-t$ into $\ep+\om$ and $\ep$, respectively,
(i.e., if we multiply $\exp[i(\ep+\om)(t-t_2)+i\ep(t_1-t)]$ and
integrate with respect to $t-t_1$ and $t_1-t$), this equation becomes
\begin{eqnarray}
&& \delta(\br_1-\br)G^{-1}(\br_1,\br_2;\ep+\om) - G^{-1}(\br_1,\br_2;\ep)\delta(\br-\br_2)
\nonumber\\
&=& i\nabla\cdot\GGamma(\br_1,\br_2,\br;\ep+\om,\ep)
+\om\Gamma(\br_1,\br_2,\br;\ep+\om,\ep).
\label{dG^{-1}-G^{-1}d=-qG+wG}
\end{eqnarray}
At the lowest order, the first term in the right hand side is approximately given by
\begin{eqnarray}
&& i\nabla\cdot\GGamma(\br_1,\br_2,\br;\ep+\om,\ep)
\nonumber\\
&\sim& \frac{1}{2}\nabla\cdot(\nabla_1+\nabla_2)\delta(\br_1-\br)\delta(\br-\br_2)
\nonumber\\
&=& - \delta(\br_1-\br)\frac{1}{2}\nabla^2\delta(\br-\br_2)
+ \frac{1}{2}\nabla_1^2\delta(\br_1-\br)\delta(\br-\br_2)
\nonumber\\
&& - i\int_{-\infty}^{\infty}\frac{d\om'}{2\pi}
G(\br_1,\br;\ep-\om')\left(-\frac{1}{2}\nabla^2\right)
G(\br,\br_2;\ep+\om-\om')W(\br_1,\br_2;\om')
\nonumber\\
&& + i\int_{-\infty}^{\infty}\frac{d\om'}{2\pi}
\left(-\frac{1}{2}\nabla_1^2\right)G(\br_1,\br;\ep-\om')
G(\br,\br_2;\ep+\om-\om')W(\br_1,\br_2,\om').
\label{ivG=-dvd+vdd}
\end{eqnarray}
Substituting this into (\ref{dG^{-1}-G^{-1}d=-qG+wG}),
we have
\begin{eqnarray}
&& \delta(\br_1-\br)G^{-1}(\br_1,\br_2;\ep+\om)
- G^{-1}(\br_1,\br_2;\ep)\delta(\br-\br_2)
\nonumber\\
&=& - \delta(\br_1-\br)\frac{1}{2}\nabla^2\delta(\br-\br_2)
+ \frac{1}{2}\nabla_1^2\delta(\br_1-\br)\delta(\br-\br_2)
\nonumber\\
&& - i\int_{-\infty}^{\infty}\frac{d\om'}{2\pi}
G(\br_1,\br;\ep-\om')\left(-\frac{1}{2}\nabla^2\right)
G(\br,\br_2;\ep+\om-\om')W(\br_1,\br_2;\om')
\nonumber\\
&& + i\int_{-\infty}^{\infty}\frac{d\om'}{2\pi}
\left(-\frac{1}{2}\nabla_1^2\right)G(\br_1,\br;\ep-\om')
G(\br,\br_2;\ep+\om-\om')W(\br_1,\br_2;\om')
\nonumber\\
&& + \omega\Gamma(\br_1,\br_2,\br;\ep+\om,\ep),
\label{d(G^{-1}-G^{-1})=-dvd+vdd+wG}
\end{eqnarray}
which is eqivalent to
\begin{eqnarray}
&& \Sigma(\br_1,\br_2;\ep)\delta(\br-\br_2)
-\delta(\br_1-\br)\Sigma(\br_1,\br_2;\ep+\om)
\nonumber\\
&=& - i\int_{-\infty}^{\infty}\frac{d\om'}{2\pi}
G(\br_1,\br;\ep-\om')\left(-\frac{1}{2}\nabla^2\right)
G(\br,\br_2;\ep+\om-\om')W(\br_1,\br_2;\om')
\nonumber\\
&& + i\int_{-\infty}^{\infty}\frac{d\om'}{2\pi}
\left(-\frac{1}{2}\nabla_1^2\right)G(\br_1,\br;\ep-\om')
G(\br,\br_2;\ep+\om-\om')W(\br_1,\br_2;\om')
\nonumber\\
&& + \omega[\Gamma(\br_1,\br_2,\br;\ep+\om,\ep)-\delta(\br_1-\br)\delta(\br-\br_2)].
\end{eqnarray}
In the linarized formulation, this equation can be rewritten as
\begin{eqnarray}
&& \Sigma(\br_1,\br_2;\ep)\delta(\br-\br_2)
-\delta(\br_1-\br)\Sigma(\br_1,\br_2;\ep+\om)
\nonumber\\
&=& - i\int_{-\infty}^{\infty}\frac{d\om'}{2\pi}
\tilde{G}(\br_1,\br;\ep-\om')\left(-\frac{1}{2}\nabla^2\right)
\tilde{G}(\br,\br_2;\ep+\om-\om')W(\br_1,\br_2;\om')
\nonumber\\
&& + i\int_{-\infty}^{\infty}\frac{d\om'}{2\pi}
\left(-\frac{1}{2}\nabla_1^2\right)\tilde{G}(\br_1,\br;\ep-\om')
\tilde{G}(\br,\br_2;\ep+\om-\om')W(\br_1,\br_2;\om')
\nonumber\\
&& + \omega[\Gamma(\br_1,\br_2,\br;\ep+\om,\ep)-\delta(\br_1-\br)\delta(\br-\br_2)].
\label{d(-S+S)=w(G-1)}
\end{eqnarray}
In order to prove this equality, we calculate the left hand side of (\ref{d(-S+S)=w(G-1)})
within the linearized formulation as follows:
\begin{eqnarray}
&& i\int_{-\infty}^{\infty}\frac{d\om'}{2\pi}
[\tilde{G}(\br_1,\br_2;\ep-\om')\delta(\br-\br_2)
-\delta(\br_1-\br)\tilde{G}(\br_1,\br_2;\ep+\om-\om')]
W(\br_1,\br_2;\om')
\nonumber\\
&=& i\int_{-\infty}^{\infty}\frac{d\om'}{2\pi}
\left[\langle\br_1|\frac{1}{\ep-\om'-\tilde{H}-i\delta_{\tilde{H}}}
|\br\rangle\langle\br|\br_2\rangle
-\langle\br_1|\br\rangle\langle\br|
\frac{1}{\ep+\om-\om'-\tilde{H}-i\delta_{\tilde{H}}}|\br_2\rangle\right]
W(\br_1,\br_2;\om')
\nonumber\\
&=& i\int_{-\infty}^{\infty}\frac{d\om'}{2\pi}
\langle\br_1|\frac{1}{\ep-\om'-\tilde{H}-i\delta_{\tilde{H}}}|\br\rangle
\langle\br|\frac{\ep+\om-\om'-\tilde{H}-i\delta_{\tilde{H}}}
{\ep+\om-\om'-\tilde{H}-i\delta_{\tilde{H}}}
|\br_2\rangle
W(\br_1,\br_2;\om')
\nonumber\\
&-& i\int_{-\infty}^{\infty}\frac{d\om'}{2\pi}
\langle\br_1|\frac{\ep-\om'-\tilde{H}-i\delta_{\tilde{H}}}
{\ep-\om'-\tilde{H}-i\delta_{\tilde{H}}}|\br\rangle
\langle\br|\frac{1}{\ep+\om-\om'-\tilde{H}-i\delta_{\tilde{H}}}|\br_2\rangle
W(\br_1,\br_2;\om')
\nonumber\\
&=& i\int_{-\infty}^{\infty}\frac{d\om'}{2\pi}
\langle\br_1|\frac{1}{\ep-\om'-\tilde{H}-i\delta_{\tilde{H}}}|\br\rangle
\langle\br|\frac{\om}{\ep+\om-\om'-\tilde{H}-i\delta_{\tilde{H}}}|\br_2\rangle
W(\br_1,\br_2;\om')
\nonumber\\
&-& i\int_{-\infty}^{\infty}\frac{d\om'}{2\pi}
\langle\br_1|\frac{1}{\ep-\om'-\tilde{H}-i\delta_{\tilde{H}}}|\br\rangle
\langle\br|\tilde{H}\frac{1}{\ep+\om-\om'-\tilde{H}-i\delta_{\tilde{H}}}|\br_2\rangle
W(\br_1,\br_2;\om')
\nonumber\\
&+& i\int_{-\infty}^{\infty}\frac{d\om'}{2\pi}
\langle\br_1|\tilde{H}\frac{1}{\ep-\om'-\tilde{H}-i\delta_{\tilde{H}}}|\br\rangle
\langle\br|\frac{1}{\ep+\om-\om'-\tilde{H}-i\delta_{\tilde{H}}}|\br_2\rangle
W(\br_1,\br_2;\om')
\nonumber\\
&=& i\om\int_{-\infty}^{\infty}\frac{d\om'}{2\pi}
\tilde{G}(\br_1,\br;\ep-\om')\tilde{G}(\br,\br_2;\ep+\om-\om')
W(\br_1,\br_2;\om')
\nonumber\\
&-& i\int_{-\infty}^{\infty}\frac{d\om'}{2\pi}
\tilde{G}(\br_1,\br;\ep-\om')\left(-\frac{1}{2}\nabla^2\right)
\tilde{G}(\br,\br_2;\ep+\om-\om')W(\br_1,\br_2;\om')
\nonumber\\
&+& i\int_{-\infty}^{\infty}\frac{d\om'}{2\pi}
\left(-\frac{1}{2}\nabla_1^2\right)\tilde{G}(\br_1,\br;\ep-\om')
\tilde{G}(\br,\br_2;\ep+\om-\om')W(\br_1,\br_2;\om')
\nonumber\\
&-& i\int_{-\infty}^{\infty}\frac{d\om'}{2\pi}\int d\br'\;
\tilde{G}(\br_1,\br;\ep-\om')\Sigma_{\rm xc}(\br,\br')
\tilde{G}(\br',\br_2;\ep+\om-\om')W(\br_1,\br_2;\om')
\nonumber\\
&+& i\int_{-\infty}^{\infty}\frac{d\om'}{2\pi}\int d\br'\;
\Sigma_{\rm xc}(\br_1,\br')\tilde{G}(\br',\br;\ep-\om')
\tilde{G}(\br,\br_2;\ep+\om-\om')W(\br_1,\br_2;\om'),
\label{wGGW}
\end{eqnarray}
where we used the fact that the electron-nucleus potential in $\tilde{H}$
commutes with the operator $|\br\rangle\langle\br|$.
The first term of Eq. (\ref{wGGW}) is exactly equal to the vertex part
at the lowest order in the dynamically screened Coulomb interaction $W$
shown in Fig. 2(b) except for the prefactor $\om$, and thus equals to
the last two terms of the right hand side of Eq. (\ref{d(-S+S)=w(G-1)}).
The second and the third terms equal to the first and second terms of
the right hand side of Eq. (\ref{d(-S+S)=w(G-1)}).
The fourth and fifth terms are at least one order higher in $v$
compared to the other terms that are lowest order in $W$,
and can be ignored.
Therefore, the LGW$\Gamma_W$ method satisfies the Ward--Takahashi identity
to the lowest order in $W$.
This discussion holds also in the case we expand to the first order
in the electron-electron Coulomb interaction $v$ of Fig. 2(a),
since $W(\br_1,\br_2;\om')$ can be replaced by $v(\br_1-\br_2)$
in this case.
Therefore, the LGW$\Gamma_v$ method also satisfies the Ward--Takahashi identity
to the lowest order in $v$.

\section{Polarization Function and Self-Energy}

The contributions to the polarization function and self-energy coming from the first-order vertex correction,
FIGs. 2(e) and (h) are explicitly given by
\begin{align}
	\tilde{P}_{\Gs\Gs'}^{(1)}(\omega) &= 
        \frac{1}{\Omega} \int \frac{d\omega'}{2\pi} e^{i\omega'\eta}
         \int \frac{d\omega''}{2\pi} e^{i\omega''\eta}
          \sum_{\s} \sum_{ijkl}\sum_{\Gs''\Gs'''} 
        W_{\Gs''\Gs'''}(\omega'')\notag\\
	&\times \frac{ \bra{\tilde{i\s}} e^{-i\Gs\cdot\rs} \ket{\tilde{j\s}} \bra{\tilde{j\s}} e^{-i\Gs''\cdot\rs''} \ket{\tilde{k\s}} \bra{\tilde{k\s}}
	 e^{i\Gs'\cdot\rs'} \ket{\tilde{l\s}} \bra{\tilde{l\s}} e^{i\Gs'''\cdot\rs'''} \ket{\tilde{i\s}}}
        {(\omega'-\ep_{i\s}-i\eta_{i\s})(\omega+\omega'-\ep_{j\s}-i\eta_{j\s})
        (\omega'+\omega''+\omega-\ep_{k\s}-i\eta_{k\s})(\omega'+\omega''-\ep_{l\s}-i\eta_{l\s})},
        \label{P1}
\end{align}
and
\vskip-5mm
\begin{align}
	\qquad 
	\Bra{\alpha}\Sigma^{(2)}_{\s}(\om)\Ket{\beta} &= 
	\int \frac{d\omega'}{2\pi} e^{i\omega'\eta}
	 \int \frac{d\omega''}{2\pi} e^{i\omega''\eta}
	  \sum_{\s} \sum_{ijk}\sum_{\Gs\Gs'}\sum_{\Gs''\Gs'''} 
        W_{\Gs\Gs'}(\omega') W_{\Gs''\Gs'''}(\omega'')\notag\\
	&\times \frac{\Bra{\alpha} e^{i\Gs\cdot\rs} \ket{\tilde{i\s}} \bra{\tilde{i\s}} e^{i\Gs'\cdot\rs'} \ket{\tilde{j\s}} \bra{\tilde{j\s}}
	e^{-i\Gs''\cdot\rs''} \ket{\tilde{k\s}} \bra{\tilde{k\s}} e^{-i\Gs'''\cdot\rs'''} \Ket{\beta}}
	{(\omega+\omega'-\ep_{i\s}-i\eta_{i\s})(\omega+\omega'+\omega''-\ep_{j\s}-i\eta_{j\s})(\omega+\omega''-\ep_{k\s}-i\eta_{k\s})},
	\label{Sigma_2c}
\end{align}
respectively,
where $W_{\bG\bG'}(\omega)$ is the dynamically screened Coulomb interaction,
$\eta$ is a positive infinitesimal ($\eta_{i\s}$ is $\eta$ for occupied $i\s$ and
$-\eta$ for empty $i\s$), $\Om$ is the volume of the unit cell,
and $\Bra{\alpha}$, $\Ket{\beta}$ denote basis functions.

\section{Interaction kernel $\tilde{I}^{\;\s_1\s'_1}_{\nu\mu dc}(\om)$}

As written in the Letter, the interaction kernel $\tilde{I}^{\sigma_3\sigma'_3}(x_3, x'_3; x_4, x'_4)$
is expressed as
$i \{ \delta(x_3-x_4)\delta(x_3'-x_4')W(x_3,x_3')
+ \delta(x_3-x_4)[W\Gamma_{v\sigma'_3}](x_3',x_4';x_3)
 + \delta(x_3'-x_4')[W\Gamma_{v\sigma_3}](x_3,x_4;x_3') \}$,
which are diagrammatically represented by
Figs. 2(i), (j), and (k), respectively.
In the third term corresponding to Fig. 2(k), for example, the two points connected by the dotted line,
i.e., the bare Coulomb interaction, are the same time $t_3=t_4$.
On the other hand, one end of the wavy line, which is involved
in this vertex part $\Gamma_{v\sigma_3}$ is another time, say $t''$,
and the other end of the wavy line,
which is not involved in the vertex part is obviously a unique time $t'_3=t'_4$.  
Therefore, this interaction kernel $\tilde{I}^{\sigma_3\sigma'_3}(x_3, x'_3; x_4, x'_4)$
has only the time dependence through $\tau=t_3-t'_3$.
Moreover, the vertex part $\Gamma_{v\sigma_3}$ in this diagram has only
the time dependence through $\tau' = t_3-t''$, and the wavy line $W$
has only the time dependence through $\tau''=t''-t'_3 = \tau-\tau'$.
Therefore, the interaction kernel $\tilde{I}^{\sigma_3\sigma'_3}(\tau)$, i.e., the dynamically screened Coulomb interaction $W(\tau-\tau')$
times the vertex correction $\Gamma_{v\sigma_3}(\tau')$, has a convolution type in time,
so that its Fourier transformation from $\tau$ to $\omega'$ is written
as $\tilde{I}^{\sigma_3\sigma'_3}(\omega')$, which can be written as the product of the Fourier transforms
of $W(\tau'')$ and $\Gamma_{v\sigma_3}(\tau')$,
$W(\omega')$ and $\Gamma_{v\sigma_3}(\omega')$;
thus we have $\tilde{I}^{\sigma_3\sigma'_3}(\omega')=W(\omega')\Gamma_{v\sigma_3}(\omega')$.
Since the $\omega'$-dependence in $\tilde{I}^{\sigma_3\sigma'_3}(\omega')$ can be treated in a standard way
given first by Strinati [S1] when we solve the Bethe--Salpeter equation,
it is allowed to replace $\tilde{I}^{\sigma_3\sigma'_3}(\omega')$ with
$W(\omega')\Gamma_{v\sigma_3}(\omega')$
in the final equation (2.18) of Ref. [S1], and thus we can derive
\begin{gather}
\hskip2mm
	\tilde{I}^{\;\s_1\s'_1}_{\nu\mu dc}(\om)
	= i \sum_{\Gs\Gs'} \int \frac{d\om'}{2\pi} e^{-i\om'\eta}W_{\Gs\Gs'}(\om')
        	\Biggl\{ \bra{\tilde{\nu\s_1}} e^{i\Gs\cdot\rs} \ket{c\s_1}
        	\bra{d\s'_1} e^{-i\Gs'\cdot\rs'} \ket{\tilde{\mu\s'_1}}
	 \notag\\
	 \qquad\qquad
         + \sum_{\alpha \beta}
		\bigl( \delta_{\alpha\, {\rm occ}} \delta_{\beta\, {\rm emp}} - \delta_{\alpha\,
 {\rm emp}} \delta_{\beta\, {\rm occ}} \bigr)
        	\Biggl( \bra{\tilde{\nu\s_1}} e^{i\Gs\cdot\rs} \ket{c\s_1}
 V_{d\alpha\beta\mu}^{\s_1'\s_1'}
		\frac{ \bra{\tilde{\alpha\s'_1}} e^{-i\Gs'\cdot\rs'} \ket{\tilde{\beta\s'_1}}}
		{\epsilon_{\alpha\sigma'_1}-\epsilon_{\beta\sigma'_1}+\om'+i\eta_{\alpha\s'_1}}
        	\notag\\
	 \qquad\qquad
         + V_{\nu\alpha\beta c}^{\s_1\s_1}\frac{\bra{\tilde{\alpha\s_1}}
       e^{i\Gs\cdot\rs} \ket{\tilde{\beta\s_1}}}
		{\epsilon_{\alpha\sigma_1}-\epsilon_{\beta\sigma_1}-\om'+i\eta_{\alpha\s_1}}
                \bra{d\s'_1} e^{-i\Gs'\cdot\rs'} \ket{\tilde{\mu\s'_1}} \Biggr) \Biggr\}
         \notag\\
                \qquad\qquad\hskip2mm
          \times
          \Biggl\{ \frac{1}{\om-\om'-\epsilon_{c\sigma_1}+\epsilon_{\mu\sigma'_1}+i\eta}
		+\frac{1}{\om+\om'-\epsilon_{\nu\sigma_1}+\epsilon_{d\sigma'_1}+i\eta} \Biggr\}
\label{BSE-0}
\end{gather}
with $V_{\nu\mu dc}^{\s_1\s_2} = \sum_{\Gs} \bra{\tilde{\nu\s_1}} e^{i\Gs\cdot\rs}
\ket{\tilde{\mu\s_1}} \bra{d\s_2} e^{-i\Gs\cdot\rs'} \ket{c\s_2} v(\bG)$
(note that all states are renormalized QP states except for $d$ and $c$).

\vskip5mm

\noindent
Reference:

\noindent
[S1] G. Strinati, Phys. Rev. B {\bf 29}, 5718 (1984).

\end{document}